\documentclass[preprint2]{aastex}

\shorttitle{Spectrophotometry with Hectospec}
\shortauthors{Fabricant et al.}

\begin{document}


\title{Spectrophotometry with Hectospec, \\
       the MMT's Fiber-Fed Spectrograph}


\author{Daniel G. Fabricant, Michael J. Kurtz, Margaret J. Geller, \\ Nelson Caldwell, and Deborah Woods}
\affil{Center for Astrophysics, Cambridge, MA 02138}

\and

\author{Ian Dell'Antonio}
\affil{Brown University, Providence, RI, 02912}

\begin{abstract}

The spectrophotometric calibration of surveys is a significant, but often neglected, issue when measuring the history of star formation by combining spectroscopic surveys conducted with different instruments.
We describe techniques for photometric calibration of
optical spectra obtained with the MMT's fiber-fed spectrograph,
Hectospec.  The atmospheric dispersion compensation prisms built
into the MMT's f/5 wide field corrector effectively eliminate errors
due to differential refraction, and simplify the calibration
procedure.  The procedures that we describe here are applicable to all 220,000+ spectra obtained to date with Hectospec because the instrument response is stable.  We estimate the internal error in the Hectospec measurements by comparing duplicate measurements of $\sim$1500 galaxies.  For a sample of 400 galaxies in the Smithsonian Hectospec Lensing Survey (SHELS) with a median z=0.10, we compare line and continuum fluxes measured by Hectospec through a 1.5$^{\prime\prime}$ diameter optical fiber with those measured by the Sloan Digital Sky Survey (SDSS) through a 3$^{\prime\prime}$ diameter optical fiber. Agreement of the [OII] and H${\alpha}$ SHELS and SDSS line fluxes,
after scaling by the R band flux in the different apertures,
suggests that the spatial variation in star formation rates over a 1.5 to 3 kpc radial scale is small.  The median ratio of the Hectospec and SDSS spectra, smoothed over 100 ${\rm \AA}$ scales, is remarkably constant to $\sim$5\% over the range of 3850 to 8000 ${\rm \AA}$. Offsets in the ratio of the median [OII] and H$\alpha$ fluxes,  the equivalent width of H$\delta$ and the continuum index d4000 are a few percent, small compared with other sources of scatter.  We also explore the  impact of atmospheric absorption.  Observing redwards of 6500 ${\rm \AA}$, it is impossible to remove the  effects of atmospheric absorption perfectly because the variation of absorption with wavelength is not resolved by a moderate dispersion spectrograph.  Thus  measurements of spectral line fluxes including H${\alpha}$, and derived physical quantities including star formation rates, may have sizable systematic errors where the redshifted spectral features land on strong atmospheric absorption troughs.

\end{abstract}


\section{Introduction}

As spectroscopic and photometric surveys increase in quality and size,
the physical parameters which govern  galaxy evolution become increasingly accessible. A range of measures of star formation rate densities from a variety of surveys provide an impressive outline of the star formation history of the universe \citep[e.g.][]{Hopkins04, Ly07, Tresse07, Reddy08, Shioya08, Villar08}.

One of the many issues in reconstructing the star formation history
is the relative calibration of surveys which use different measures
(Hopkins 2004). After calibration, the scatter in the
measured star formation rate density at fixed redshift remains large.
Even for the same star formation indicator, inadequacy of the relative
calibration of surveys with different instruments on different telescopes may be one source of scatter and systematic offset.
Here we use the overlap of  two large spectroscopic samples from different telescopes with different fiber instruments to
evaluate these issues for emission line fluxes, H$\delta$ equivalent width, and the strength of the 4000 {\rm \AA} break. We demonstrate that with attention to technical detail, the median offsets in all of these measures are at the few percent level.

One of our overlapping samples is the SHELS survey {\citep{Geller05} carried out with the Hectospec \citep{Fabricant05} on the MMT; the other is derived from the Sloan Digital Sky Survey \citep{DR6}. We describe the spectrophotometric calibration of Hectospec and estimate the internal errors in this calibration by examining 1467 unique pairs of measurements of a subset of the galaxies.  We estimate the external
errors in the spectrophotometric calibration by comparisons with
galaxies observed in common with the Sloan Digital Sky Survey \citep[SDSS,][]{DR6}.

We extract the absolute calibration for our spectrophotometry from
external photometry; we use the Deep Lensing Survey R-band photometry
\citep{Wittman06, Wittman02}.  This absolute calibration
corrects for clouds, seeing, telescope tracking and guiding, as well
as errors in astrometry and alignment, as long as these losses of
light are independent of wavelength.  The spectrophotometric errors
associated with differential refraction \citep{Filippenko82} are largely eliminated because the MMT's wide-field corrector \citep{Fabricant04} contains atmospheric dispersion compensation (ADC) prisms.  The atmospheric dispersion across Hectospec's wavelength range ($\sim$3700 to 8500 {\rm \AA}), if not corrected by the ADC prisms,  would exceed the Hectospec's 1.5$^{\prime\prime}$ fiber diameter at a zenith angle of 45$^{\circ}$, or an airmass of 1.4.

Section 2 describes the data and the procedures  we use to calibrate the Hectospec spectrophotometry. Because Hectospec's throughput is stable (Section 2.3), this calibration procedure is applicable to all Hectospec observations. In Section 3 we measure the internal errors in the Hectospec spectrophotometry by comparing repeated observations; in Section 4 we estimate the external errors in our spectrophotometry by comparing Hectospec and SDSS spectra. In Section 5 we discuss the impact of calibration issues on determination of the star formation history. We conclude in Section 6.

\section{Spectrophotometry with Hectospec}

\subsection{Introduction}

In this section we describe the procedures  we use to calibrate
the Hectospec spectrophotometry.  A web page,
tdc-www.harvard.edu/instruments/hectospec/reduce.html, and \citet{Mink07} describe the standard CfA Hectospec reduction pipeline  (see also \citet{Fabricant05}).  We further process the spectra emerging from this pipeline to convert observed counts to flux.  There are four steps in this process: (1) we correct for the smooth component of atmospheric extinction, (2) we remove, as far as possible, the sharp H$_2$O and O$_2$ atmospheric absorption lines at wavelengths redder
than $\sim$6000 {\rm \AA}, (3) we correct for Hectospec relative
throughput as a function of wavelength, and (4) we compare a synthetic
R-band flux derived from the spectra to a corresponding R band
aperture magnitude to determine the absolute flux normalization.

\subsection{Data Sources}

To calibrate and test our spectrophotometric reductions we use data
from three surveys: the Smithsonian Hectospec Lensing Survey (SHELS;
\citet{Geller05}); the Deep Lens Survey \citep[DLS:][]{Wittman06,Wittman02}; and the Sloan Digital Sky Survey \citep[SDSS:][]{DR6}.

The SHELS F2 galaxy sample includes spectra for 9825 galaxies to a limiting total R magnitude of 20.3 in a field of 4 square degrees. The R band magnitudes are on the Cousins system, \citet{Bessell98}.  The integral SHELS completeness to the limiting magnitude is 97.7\%, and the differential completeness at the limiting magnitude is 95.5\%.  The F2 field covers the RA range 9.24667 - 9.40389 hr and the DEC range 28.97722$^{\circ}$ - 31.02333$^{\circ}$.  The spectra were obtained with Hectospec on the MMT between April 13, 2004 and April 20, 2007.

From the SHELS dataset we use 2934 SHELS spectra (1467 pairs) to calibrate our internal errors and 406 spectra for galaxies which also have SDSS spectrophotometry \citep{DR6} to calibrate our external errors.

The SHELS galaxies are selected based on DLS total R magnitudes; here we use DLS R aperture photometry to calibrate our spectra to absolute units of flux.  We use 1.5$^{\prime\prime}$ diameter aperture photometry to calibrate the flux in our 1.5$^{\prime\prime}$ fibers, and we use 3$^{\prime\prime}$ aperture magnitudes to transform our spectrophotometry to match the spectrophotometry of the SDSS, transformed to their 3$^{\prime\prime}$ fiber apertures.

The SHELS spectra were obtained with Hectospec's 270 line mm$^{-1}$ grating that provides a dispersion of 1.2 {\rm \AA} pixel$^{-1}$ and a resolution of 6.2 {\rm \AA} FWHM.  Wavelength calibration is carried out with a hollow Fe cathode NeAr lamp illuminating a screen mounted on the MMT enclosure; typical scatter in the fits of the wavelengths of line centers to low order polynomials is 0.06 {\rm \AA}. Exposures range between 45 and 120 minutes. Hectospec's 300 fibers can be placed within a circular field of view 1$^{\circ}$ in diameter.  Typically 10 to 15\% of the fibers are used for sky background measurement.  Sky subtraction accuracy is ultimately limited by the fiber to fiber throughput variations (2\% RMS) that are calibrated with dome flats or twilight flats. System throughput (Hectospec plus MMT) for light entering a 1.5$^{\prime\prime}$ aperture peaks at 17\% at $\sim$6000 {\rm \AA}, falling to half this value at 3900 and 7900 {\rm \AA}.
Full details are given in \citet{Fabricant05}.

\subsection{Atmospheric Extinction}

High precision photometry or spectrophotometry requires frequent
measurements of atmospheric extinction, but the necessary observations
are costly in telescope time.  Instead, we use a standard
extinction curve obtained at KPNO and distributed with IRAF.  We do not have a measure of the night to night extinction variations, and such variation is included when we measure internal or external errors in our measurements in Sections 3 and 4.  The standard
extinction curve only includes the continuum extinction and not the
narrow H$_2$O and O$_2$ absorption features increasingly
prominent in the red.  The absorption from these narrow features can
only be removed in an average sense from moderate dispersion spectra;
the intensities of spectral lines in the target object that fall
in the absorption troughs are correspondingly uncertain.

Figure~\ref{fig1} shows the narrow absorption features between 6500 and 9000 {\rm \AA} plotted at 2 {\rm \AA} resolution; we also show a spectrum of the night sky at the same resolution.  The absorption features are taken from a high dispersion atlas obtained with the Fourier Transform Spectrograph at the McMath Solar Observatory available at ftp.noao.edu/catalogs/atmospheric\_transmission/.  The upper axis shows the redshift of H${\alpha}$ at the wavelengths plotted on the lower axis. Regions free of strong night sky emission lines are not necessarily free of strong absorption features.

The McMath measurement of atmospheric absorption lines is
approximately normalized to one airmass at an altitude of 2096 meters.
At the MMT altitude of 2600 meters, the absorption is somewhat less.
We have empirically renormalized the McMath absorption to produce the
best average correction of the Hectospec spectra for the strong
absorption features between 6700 and 8000 {\rm \AA}; reducing the
McMath absorption by 15\% provides the best result.

The standard Hectospec pipeline applies a partial correction for the
atmospheric absorption features, including only the strong features at
$\sim$6800 and $\sim$7600 {\rm \AA}.  We remove this pipeline
correction and apply a correction using all of the features in the
McMath spectrum, binned to a resolution of 1 ${\rm \AA}$.  Because the
absorption varies on a wavelength scale finer than the resolution of a
moderate dispersion spectrograph, the absorption correction is
approximate.  Although the corrected continuum looks smooth, the
fluxes from narrow emission lines are not measured to correspondingly
high accuracy.

We can estimate the potential errors in line fluxes for narrow emission lines by calculating the difference between the McMath absorption at 0.1 and 2 {\rm \AA} resolution, normalized by the absorption at 2 {\rm \AA} resolution.  For regions with strong absorption, for example 6800-6900 {\rm AA} and 8100-8300 {\rm \AA}, the mean normalized difference is 7-8\%, and the RMS normalized difference is 14-17\%.

\subsection{Hectospec Relative Throughput}

We observe spectrophotometric standard stars periodically with
Hectospec to monitor its throughput. In the early operation of the instrument we allowed many months between standard star observations, although the current practice is to observe standard stars at least once per run.  Fortunately, Hectospec's long term stability has proven to be excellent as discussed below.  No secondary spectrophotometric standards were observed simultaneously with the SHELS galaxy observations \citep[e.g.][]{Stoughton02}.

One complication in interpreting
the standard star spectra is the presence of a significant amount of
second order light at wavelengths redwards of 7000 ${\rm \AA}$ when
observing blue spectra.  We do not currently use an order blocking
filter with Hectospec because the loss of spectrophotometric precision
is outweighed by the desire for maximum throughput.

For maximum spectrophotometric accuracy it is necessary to use a
standard star with a spectrum as similar as possible to the observed
objects.  Most of the observed spectrophotometric standard stars are
significantly bluer than the average galaxy spectrum in the SHELS
survey.  One standard \citep{Massey88}, VI Cyg (Cyg OB2 No.9), a reddened O5I supergiant, has a very red spectrum that reduces second order contamination from the grating to a negligible level, and is thus well matched to a typical SHELS galaxy.  We use data obtained from VI Cyg in April 2007 to calibrate the Hectospec throughput for the spectrophotometry described here.

We have spectrophotometric observations of BD+284241, a subdwarf O
star with a blue spectrum, from widely separated times.  These spectra
allow us to look for changes in instrument throughput over this time.
Figure~\ref{fig2} shows the ratio of spectra obtained in October 2004 and November 2007, normalized to unity at 6000 ${\rm \AA}$.  Differences in seeing and extinction complicate this comparison, but the relative throughput appears to be constant to $\pm$10\% between 4000 and 8000 ${\rm \AA}$ over three years.

\subsection{Absolute Spectrophotometric Calibration}

Prior to the absolute spectrophotometric calibration, the spectra
processed by the standard pipeline are corrected for extinction,
converted from counts to flux in f$_{\lambda}$ units (ergs cm$^{-2}$ sec$^{-1}$ {\rm \AA$^{-1}$), and corrected for relative throughput as a function of wavelength.  The absolute flux normalization is arbitrary because the observations were made in variable seeing and possibly through clouds.

We begin our absolute spectrophometric flux calibration by calculating a synthetic R-band flux in f$_{\lambda}$ units.

\begin{equation}
f_R={\alpha\cdot{\int{F({\lambda})T(\lambda})d{\lambda}}\over{\int{T(\lambda})d{\lambda}}}
\end{equation}

Here, $f_R$ is the flux at the R band effective wavelength ($\sim$6410 {\rm \AA} for the Cousins system, \citet{Bessell98}), $F(\lambda)$ is the raw Hectospec flux as a function of wavelength, and $\alpha$ is a normalization constant required to convert the raw Hectospec flux to a calibrated flux. All fluxes are in f$_{\lambda}$ units. We calculate $T(\lambda)$, the R-band response for the KPNO Mosaic camera system \citep{Muller98}, as the product of the K1004 filter response and the throughput of the Mosaic CCDs.   The final step is to determine the correct normalization, $\alpha$, by converting the R-band 1.5$^{\prime\prime}$ diameter aperture magnitudes to an R-band flux (ergs cm$^{-2}$ sec$^{-1}$ {\rm \AA$^{-1}$).  We use the results tabulated in Table A2 of \citet{Bessell98} (note that the f$_{\lambda}$ and f$_{\nu}$ zero points are transposed), where

\begin{equation}
f_R=10^{(-0.4*(R+21.655))}
\end{equation}

\subsection{Index Definition}

We use [OII] $\lambda$3727, H${\delta}$, and H${\alpha}$ line
fluxes and equivalent widths, and d4000, a measure of the strength of
the 4000 ${\rm \AA}$ spectral break.  For the line fluxes and
equivalent widths we use indices rather than fits to the line
profiles. For narrow lines the results are freer of systematic fitting
errors to the flat-topped point spread function arising from the
resolved fiber image.  Table 1 defines the wavelength regions used to
measure the [OII] $\lambda$3727, H${\delta}$, and H${\alpha}$ line
fluxes and equivalent widths.  Following \citet{Balogh99}, we define d4000, the strength of the 4000 ${\rm \AA}$
break, as the ratio of flux in the 4000--4100 ${\rm \AA}$ band to that
in the 3850--3950 ${\rm \AA}$ band.

\section{Hectospec Internal Spectrophotometric Errors}

We estimate the internal errors in the Hectospec
spectrophotometry from a sample of 1468 unique pairs of duplicate
SHELS spectra.  This large set of repeat measurements allows us to
estimate our systematic errors accurately. We include only spectra
obtained when the ADC prisms were operating properly (see Table
2).

We compare multiple flux measurements of [OII] $\lambda$3727 and H${\alpha}$, multiple equivalent width measurements of H${\delta}$, and multiple measurements of d4000.  We propagate the statistical errors for each pixel of the spectra for the equivalent width and line flux measurements.

Additional systematic errors arise from several sources: short term variations in the wavelength dependence and amount of extinction, the wavelength dependence of seeing variations, variable alignment of the fiber with respect to the observed galaxy, errors in our calibration of fiber to fiber throughput variations, imperfect sky subtraction, and the (small) wavelength dependence of cloud extinction.  We estimate the magnitude of these systematic errors by examining the differences between repeated measurements.  We add a systematic error in quadrature to the statistical errors to calculate a total internal error.  We determine the total internal error in each measurement, ${\sigma}_{t}$, by requiring 68\% of the measurement differences to lie within $\sqrt{2}{\sigma}_{t}$.  The systematic error can then be easily calculated.

We first compare 762 repeated measurements of the [OII] $\lambda$3727 and H${\alpha}$ emission line fluxes; these are plotted in Figure~\ref{fig3}.  We only include measurements
where the EW([OII]) or EW(H$\alpha$) $\geq$3 {\rm \AA},
and the emission line is significant at $\geq$3$\sigma$.  We determine the systematic error to be 18\% of the line flux.  The total RMS scatter in the measurements is 22\%, and 6.8\% of the measurement differences are larger than $\sqrt{2} (2 \sigma_t)$, to be compared with 4.5\% for a Gaussian error distribution.

Figure~\ref{fig4} plots 592 repeated measurements of H${\alpha}$ emission line fluxes. We determine the systematic error to be 18\% of the line flux.  The total RMS scatter in the measurements is 23\%, and 9.1\% of the measurement differences are larger than $\sqrt{2} (2 \sigma_t)$.

Figure~\ref{fig5} plots 1011 repeated measurements of H${\delta}$ equivalent width where the total measurement error (statistical and systematic) is 3 {\rm \AA} or less.  We determine the systematic error to be 2.3 times the statistical error.  The total RMS scatter in the measurements is 1.5 {\rm \AA}, and 4.4\% of the measurement differences are larger than $\sqrt{2} (2 \sigma_t)$.

Figure~\ref{fig6} plots all 1468 repeated measurements of d4000. We determine the systematic error to be 4.5\% of the index value.  The formal statistical error for this measurement is very low, so the total error is dominated by the systematic error.  The total RMS scatter in the measurements is 0.086, and 9.8\% of the measurement differences are larger than $\sqrt{2} (2 \sigma_t)$.

In all cases, the systematic errors dominate the statistical errors.
We note that the systematic errors in the repeated measurements of [OII] and H$\alpha$ line fluxes are both 18\% even though these lines are widely separated in observed wavelength.  The d4000 measurements appear quite robust and their errors are considerably smaller than the line flux errors.  The internal error estimates are summarized in Table 3.

\section{Comparison with SDSS Spectra}

To evaluate the external errors in our measurements of emission line
fluxes, equivalent widths, d4000, and continuum shapes, we compare 406
galaxies within both the SHELS sample and the well-calibrated Sloan Digital Sky Survey (SDSS) Data Release 6 (DR6). The set of overlapping spectra includes galaxies from the LRG survey \citep{Eisenstein01} (~25\%) with the remainder from the main sample of SDSS galaxies \citep{Strauss02}.

We reanalyze the fluxed DR6 SDSS spectra in the same fashion as the SHELS spectra to ensure a precise comparison with the SHELS results.
The DR6 and the SHELS flux calibrations are independent of one another. We do not use the published SDSS measurements of line fluxes or spectral indices because our goal is to compare the flux spectra, not analysis techniques. The median redshift of the SHELS/SDSS sample is 0.13 for the full sample of 406 galaxies; the median redshift is 0.10 for the 154 galaxies with H${\alpha}$ emission.  Figure~\ref{fig7} is the redshift histogram for the galaxies with H${\alpha}$ emission, and for the full sample.

A major subtlety in comparing the SHELS and SDSS spectra is the difference in fiber apertures, 1.5$^{\prime\prime}$ and 3.0$^{\prime\prime}$, respectively. We finesse this problem by using the DLS photometry taken with seeing between 0.82 and 0.91$^{\prime\prime}$.  The DLS photometry provides the 1.5$^{\prime\prime}$ and 3.0$^{\prime\prime}$ aperture R-band magnitudes we use for relative calibration of the SHELS and SDSS spectra; we scale the fluxes through the Hectospec and SDSS apertures to these R-band magnitudes. This procedure assumes that the emission line fluxes scale as the R-band light.  Physical scatter in this comparison certainly results from variation in the distribution of emission-line regions relative to the continuum emission in the galaxies.

One potential confusion in comparing the SHELS and SDSS spectrophotometry
is that the SDSS DR6 spectra are flux calibrated to the
integrated light from standard stars, not the portion of the light
that arrives within a fiber aperture \citep[see][]{DR6}.  This procedure might be convenient for stars, but is less so for extended objects like galaxies.  We have looked at a selection of SDSS stars in the SHELS field to derive a correction factor; the total (PSF) magnitudes are 0.34 magnitudes brighter than the 3.0$^{\prime\prime}$ aperture magnitudes.  This result agrees with the mean difference reported in \citet{DR6}.

The seeing at the time of the SHELS Hectospec observations is not a factor in the comparison between the SHELS and SDSS spectra because the SHELS spectrophotometry is normalized to the DLS photometry.  In fact, the normalization of the flux entering the Hectospec 1.5$^{\prime\prime}$ aperture drops out of the SHELS-SDSS comparison because we use the R-band photometry to convert the SHELS flux to a 3.0$^{\prime\prime}$ aperture.

Figure~\ref{fig8} plots the median ratio of 273 SHELS and SDSS galaxy spectra between 3850 and 8200 {\rm \AA}, boxcar smoothed over 100 {\rm \AA} intervals. All pairs of spectra with at least 25 counts per pixel in the raw SHELS spectra over the entire spectral range are included in the median.  The ratios of the spectra are normalized to unity between
5000 and 5500 {\rm \AA}.  The interquartile range is also plotted.  On
average, the SHELS fluxing procedure preserves the spectral
shape to high accuracy, on the order of $\pm$5\%.

The comparison of [OII] $\lambda$3727 and H${\alpha}$ line fluxes
measured from the SHELS and SDSS spectra shows that the SHELS flux calibration procedures work remarkably well.  Figure~\ref{fig9} plots SDSS [OII] lines fluxes against SHELS line fluxes for 128 galaxies with [OII] equivalent width $\geq$3 {\rm \AA} and are significant at $\geq$3 $\sigma$ confidence.  The ratio of the SHELS median flux to the SDSS median flux is 0.96.  We calculate a systematic error for the SDSS line fluxes and equivalent widths in same fashion as we did above for the SHELS systematic errors, but this time using the total errors determined above for the SHELS data.  The assignment of this additional systematic to the SDSS data is not above question, but we believe that systematic scatter affecting the SHELS data should have shown up in our internal comparisons.  The statistical errors for the SDSS measurements are propagated from the errors associated with each pixel.

We determine a systematic error of 12\% for the SDSS [OII] flux measurements.  The total RMS scatter in the [OII] flux measurements is 22\% (assuming an equal division of errors between SDSS and SHELS; the RMS scatter in the differences between the measurements is $\sqrt{2}$ larger).  The fraction of $>$2$\sigma$ outliers is 4.7\%.

Figure~\ref{fig10} is the corresponding plot for the H${\alpha}$ line fluxes of 154 galaxies where the line has an equivalent width $\geq$3 {\rm \AA} and is significant at $\geq$3 $\sigma$. Here, the ratio of median SHELS to SDSS line flux is 1.03, and an additional systematic error of 15\% is determined for the SDSS line fluxes.  The total RMS scatter for measurements is 22\%, and the fraction of $>$2$\sigma$ outliers is 9.1\%.

Figure~\ref{fig11} compares the H${\delta}$ equivalent widths measured from the SHELS and SDSS spectra.  We plot data for 352 galaxies with total measurement errors (statistical and systematic) of 3 {\rm \AA} or less.  The ratio of the SHELS to the SDSS median equivalent width is 0.97. We calculate that the SDSS total error is 1.2 times the statistical error.  The total RMS scatter per measurement is 1.0 {\rm \AA}, and the fraction of $>$2$\sigma$ outliers is 5.1\%.

Figure~\ref{fig12} compares the d4000 measured from SDSS and SHELS spectra for 358 galaxies.  The ratio of the median SHELS to the median SDSS d4000 measurement is 1.0.  We determine that the SDSS total error in d4000 is 0.03 times the index value. The total RMS scatter per measurement is 0.064, and the fraction of $>$2$\sigma$ outliers is 5.6\%.

Table 4 summarizes the comparison of the SHELS and SDSS measurements.
Our comparisons of emission line fluxes and continuum features in
spectra of the same objects taken from the SDSS DR6 and SHELS show that spectrophotometric measurements from these two fiber surveys agree remarkably well; there are no significant offsets. The calibration
procedures for the two instruments differ but yield consistent results.

The errors in emission-line fluxes substantially exceed errors in the continuum measure, d4000. This difference results in part from both the discreteness of the emission-line regions and the variation
in their distribution relative to the continuum light we use to normalize the SHELS spectra. The very small errors in d4000 and the absence of any detectable variation relative to the larger aperture SDSS photometry suggest that the underlying stellar population has little variation over the radial scales, 1.4 and 2.8 kpc (at z=0.1), sampled by the apertures.  The small d4000 measurement errors and its comparative immunity to aperture effects encourage its use as a robust measure of the star formation history \citep{Kauffmann03}.

\section{Astrophysical Implications}

\subsection{Introduction}

Understanding star formation in galaxies and galaxy evolution requires spectrophotometry \citep[e.g.][]{Kennicutt92a, Kennicutt92b, Gavazzi02, Shioya02, Lamareille05, Lamareille06, Moustakas06a, Moustakas06b, Weiner07, Cooper08}.  Star formation rates and reddening estimates require line flux measurements. Exploring these astrophysical issues across a large redshift range generally requires a combination of surveys from different telescopes. These comparisons may be compromised without a demonstration that the surveys are properly calibrated with respect to one another.

With attention to a range of technical details, fiber spectroscopy provides robust line fluxes. In the case of Hectospec, spectrophotometry in the fiber aperture is possible because the
instrument includes ADC prisms. By combining SDSS spectroscopy
with Hectospec data, many galaxy evolution issues can be profitably explored from redshift zero to 0.7 or more; for quasar spectroscopy the redshift range is much larger. As examples of the possible  impact of combining such surveys, we comment on the impact of aperture effects and on some issues in the determination of star formation rate densities.

\subsection {Aperture Effects}

Aperture effects are an important limitation in comparing samples of galaxies over a broad redshift range.  \citet{Kewley05} discuss the complex range of issues which result from a spectroscopic aperture of fixed angular size rather than a fixed physical size. By comparing nuclear spectroscopy with integrated spectroscopy of the Nearby Field Galaxy Sample \citep[NFGS:][]{Jansen00}, they demonstrate that apertures enclosing 20\% or more of the galaxy light provide an unbiased view of the star formation properties. They emphasize that the physical aperture which includes 20\% of the light is a function not only of galaxy type but also of galaxy luminosity. Thus at greater redshift where surveys are limited to the intrinsically most luminous objects, a larger physical aperture is necessary.

Figures 6 and 7 of \citet{Kewley05} are useful for judging the adequacy of the Hectospec and SDSS apertures as a function of redshift and galaxy properties. They show that for the mean NFGS galaxy, the SDSS fibers include more than 20\% of the light at z$>$0.05; the Hectospec fibers cover this fraction at z$>$0.1. For galaxies brighter than the characteristic luminosity in the \citet{Schechter76} parameterization, L$^*$,  the SDSS fibers include at least 20\% of the light only for z$>$0.1 and the Hectospec correspondingly at z$>$0.2.

The good agreement of the SHELS and SDSS [OII] $\lambda$3727 and
H${\alpha}$ line fluxes indicates that star formation rates within
apertures of 1.5$^{\prime\prime}$ and 3$^{\prime\prime}$ scale with the R-band light.  The median line fluxes agree within 3-4\% and, for high signal-to-noise spectra, the scatter is dominated by systematics which imply a $\sim$18\% typical error.

At z=0.1, the Hectospec and SDSS apertures correspond to radii of
1.4 and 2.8 kpc, respectively. The median 25\% light radius for the sample galaxies is 2.1 kpc. Thus the SDSS aperture generally includes 20\% of the light, but the Hectospec aperture includes less than the \citet{Kewley05} fiducial fraction.

Although the median luminosity of the sample galaxies is
R = -21.5, approximately L$^*$ (Brown et al. 2001), and the apertures are small, aperture effects appear negligible in this dataset.  Figure~\ref{fig13} shows the ratio of H$\alpha$ fluxes for the SDSS and SHELS as a function of the absolute magnitude of the objects. Even though the SHELS fibers generally include less than 20\% of the light,
there is no apparent dependence of the flux ratio on the absolute magnitude.  We have also checked that the flux ratio is independent of redshift, apparent magnitude, and 25\% light radius.

\subsection{Star Formation Rates, Reddening, and  Star Formation History}

Star formation rate indicators are a cornerstone of studies of galaxy
evolution. We focus on issues underlying the use of optical spectroscopic indicators H$\alpha$ and [OII].

A variety of star formation studies use equivalent widths of these lines as star formation indicators \citep{Poggianti99, Mouchine05}. A drawback of this procedure is the lack of reddening and metallicity measurements that require line fluxes. \citet{Kewley04} discuss some of the pitfalls of assuming a reddening independent of galaxy properties and redshift.  An additional issue not widely discussed in the literature is the systematic impact of the atmospheric absorption features on H$\alpha$ (see Figure~\ref{fig1}); these effects are a complex function of redshift.

Measures of reddening provided by spectroscopy are
also important for surveys covering a large redshift range.
In general, surveys include intrinsically more luminous objects at
greater redshift; these objects tend to be dustier and more
metal rich \citep{Calzetti01}. Thus the common assumptions of fixed reddening and metallicity can bias our view of the star formation history. Our analysis indicates that for high signal-to-noise spectra
where the systematic error dominates, the 1$\sigma$ error in  H$\alpha$/H$\beta$ is $\sim$25\%.

Strong atmospheric absorption is an additional serious issue at redshifts near 0.05, 0.1, 0.16, and 0.25. The fact that the SHELS and
SDSS H${\alpha}$ line fluxes agree well does not demonstrate that the narrow atmospheric absorption features are well corrected.  Both surveys lack the spectral resolution to remove the effect of these features with standard techniques.

Figure~\ref{fig1} shows that the corresponding wavelength ranges are relatively free of strong night sky emission, but they fall in bands of significant atmospheric absorption. A number of imaging surveys geared toward measuring the H$\alpha$ luminosity function as a route toward a star formation rate density may be biased by the impact of these absorption features. For example, \citet{Shioya08} make a state-of-the-art measurement of the H$\alpha$ luminosity function for the COSMOS survey. They use a narrow band filter centered at 8150 {\rm \AA} with a 120 {\rm \AA} width. Their survey is thus centered at $z = 0.24$.  Figure~\ref{fig1} shows that at many redshifts within the bandpass the atmospheric absorption is 20 -- 30\%. This systematic bias toward underestimating the volume averaged star formation rate is unavoidable in an imaging survey that overlaps the atmospheric absorption features. Unless the redshifts of individual objects are known from spectroscopic measurement, it is impossible to make a correction for absorption a posteriori because the impact of these features is a complex function of the redshift. Without redshifts, it is not possible to estimate overall impact of the bias. These absorption features also impact [OII] and Ly$\alpha$ imaging surveys in the same bandpass \citep{Murayama07, Takahashi07}.

We conclude that an unbiased measurement of the volume average star formation rate from an imaging survey requires knowledge of the redshifts of the objects. Spectroscopy enables an approximate  correction for the systematic underestimate of line fluxes. It also provides an independent estimate of the completeness of catalogs derived from the narrow-band images.

\section{Conclusions}

The most difficult spectrophotometric issues hampering astrophysical
measurements such as the evolution of the star formation rate density
affect all instruments, not just those with optical fibers.  These
issues include: (1) calibration across different surveys and
several instruments, (2) aperture effects on the derived star formation rate, and (3) the highly structured atmospheric absorption bands
redwards of 6000 {\rm \AA}.

Because of Hectospec's stability and the MMT's f/5 corrector with ADC prisms that eliminate errors from differential refraction, a relatively simple procedure for flux calibration of Hectospec spectra, using infrequent observations of spectrophotometric flux standards for calibration of relative throughput and reference to R-band photometry for absolute throughput, works surprisingly well. Spectrophotometric calibration of instruments without ADC prisms requires observation of secondary spectrophotometric standards in each field. The success of the procedure we describe, when tested against well-calibrated SDSS spectra, demonstrates that short term extinction variations are not a serious issue.   Over its 4.5 year operational lifetime, the Hectospec has acquired over 220,000 spectra that can be calibrated using our procedure.

We recover the SDSS continuum shapes and line fluxes to $\sim\pm$10\% for a subset of the SHELS galaxies also observed by the SDSS, allowing only a rescaling between the different aperture sizes based on R-band aperture magnitudes.  The median emission line fluxes, H$\delta$ equivalent widths, and the amplitude of the continuum break, d4000, for the SDSS and SHELS spectra agree to within a few percent.  For high signal-to-noise SHELS spectra the typical errors in emission line fluxes are 18\%.  We plan to use these measures to further explore spectroscopic investigation of the star formation history over the range of redshifts covered by the combined SDSS and SHELS surveys
\citep{Kewley09}.

\acknowledgments

We thank the Hectospec engineering team including Robert Fata, Tom Gauron, Edward Hertz, Mark Mueller, and Mark Lacasse, and the instrument specialists Perry Berlind and Michael Calkins.  We are grateful for the contributions of the members of the CfA's Telescope Data Center including Doug Mink, Susan Tokarz, and William Wyatt.  The entire staff of the MMT Observatory has done an outstanding supporting Hectospec operations.  We thank our referee, Michael Strauss, for insightful comments that helped us improve the paper.

Funding for the SDSS and SDSS-II has been provided by the Alfred P. Sloan Foundation, the Participating Institutions, the National Science Foundation, the U.S. Department of Energy, the National Aeronautics and Space Administration, the Japanese Monbukagakusho, the Max Planck Society, and the Higher Education Funding Council for England. The SDSS Web Site is http://www.sdss.org/.  The SDSS is managed by the Astrophysical Research Consortium for the Participating Institutions.

{\it Facilities:} \facility{MMT(Hectospec)}, \facility{SDSS}.

\clearpage



\begin{figure} \plotone{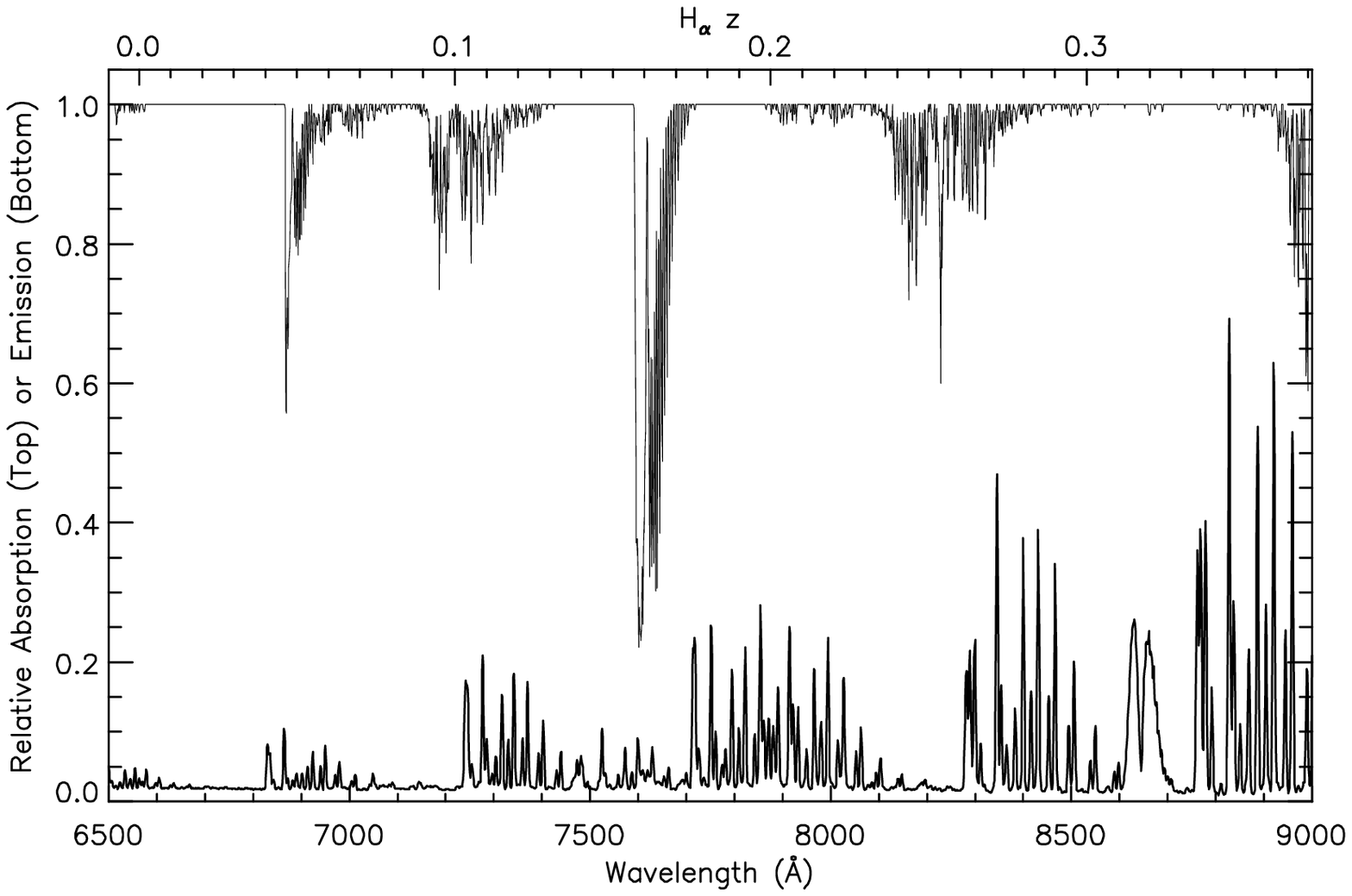}
\caption{Spectrum of sky absorption features and of the night sky plotted at 2 {\rm \AA} resolution. The night sky spectrum is in units of relative f$_{\nu}$. \label{fig1}}
\end{figure}

\begin{figure}
\plotone{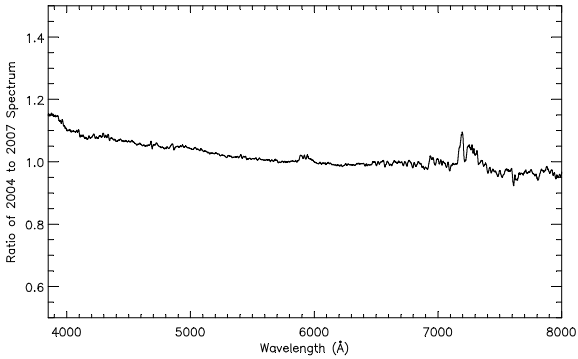}
\caption{Ratio of two
spectrophotometric standard star observations (BD+284241) in October 2004 and November 2007. \label{fig2}}
\end{figure}

\begin{figure}
\plotone{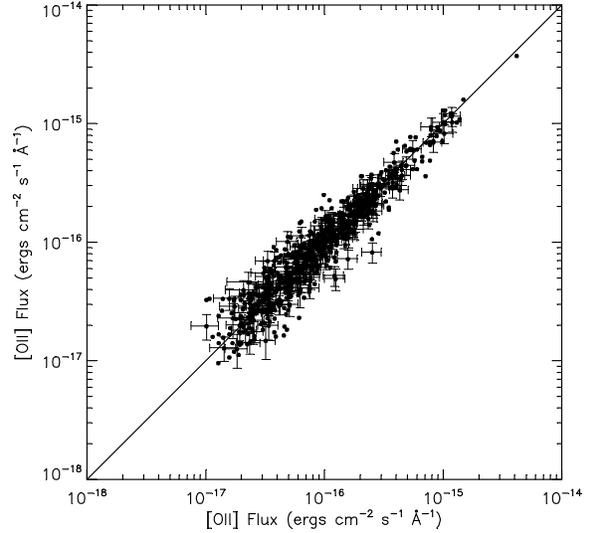}
\caption{Comparison of 762
repeated observations of [OII] $\lambda$3727 line fluxes. We include
only measurements where the EW [OII] $\geq$3 {\rm \AA} and $\geq$3
$\sigma$ significance. Representative error bars are shown. \label{fig3}}
\end{figure}

\begin{figure}
\plotone{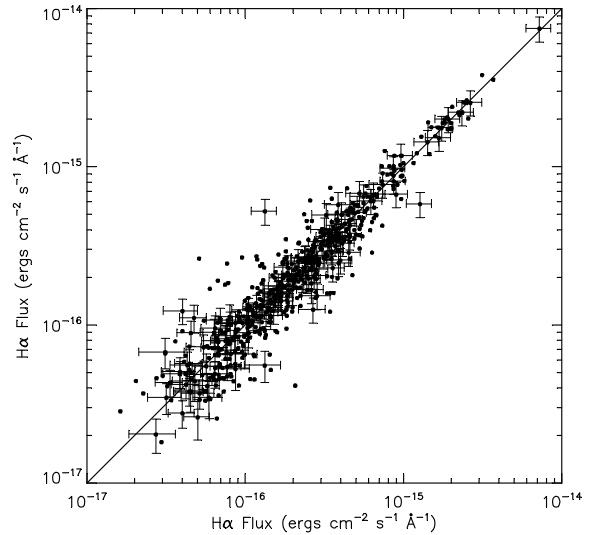}
\caption{Comparison
of 592 repeated observations of H${\alpha}$ line fluxes. We include
measurements with EW H${\alpha}$ $\geq$3 {\rm \AA} and $\geq$3
$\sigma$ significance.  \label{fig4}}
\end{figure}

\begin{figure}
\plotone{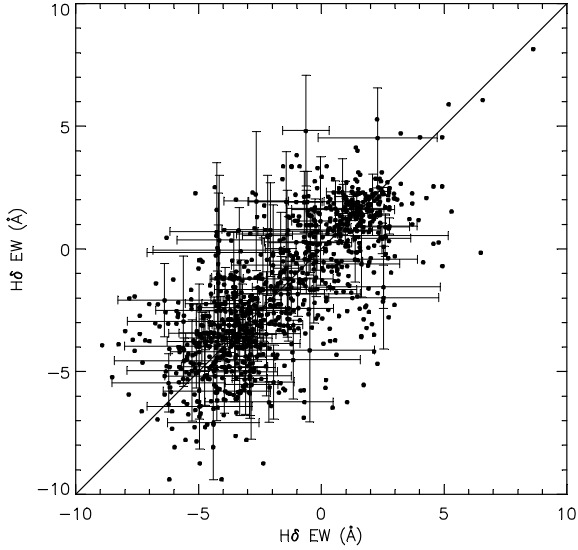}
\caption{Comparison of 1011
repeated observations of H${\delta}$ equivalent width. Measurements
with a total error $\leq$3 {\rm \AA} are included.  \label{fig5}}
\end{figure}

\begin{figure}
\plotone{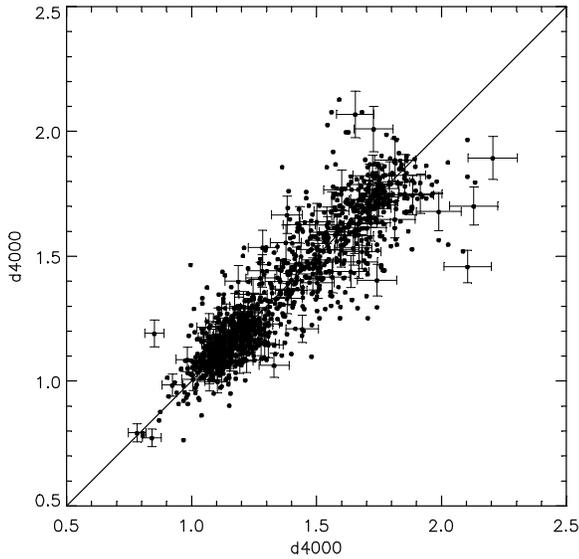}
\caption{Comparison of 1468
repeated observations of d4000, the ratio of flux in the in the
4000--4100 {\rm \AA} band to that in the 3850--3950 {\rm \AA} band. \label{fig6}}
\end{figure}

\begin{figure}
\plotone{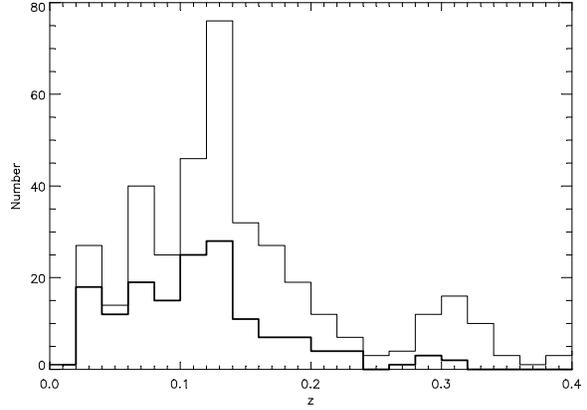}
\caption{Histogram of
redshifts in the full sample of 406 galaxies observed by both SHELS and SDSS (light line). Histogram of
redshifts for the 154 galaxies used to compare SHELS and SDSS H${\alpha}$ line fluxes (heavy line).  The median redshift for the full sample is 0.13; the median redshift for the H${\alpha}$ line flux sample is 0.10.  \label{fig7}}
\end{figure}

\begin{figure}
\plotone{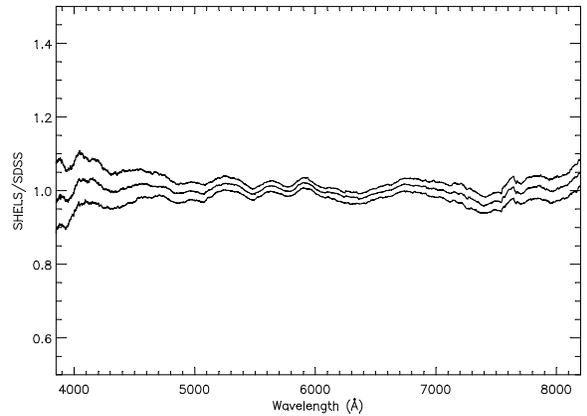}
\caption{Median ratio of
SHELS to SDSS spectra for 273 galaxies, and the interquartile range.
Only those SHELS spectra with least 25 ADU per pixel between 3850 to
8000 {\rm \AA} are included.  The ratio is normalized to one between
5000 and 5500 {\rm \AA}.  \label{fig8}}
\end{figure}

\begin{figure}
\plotone{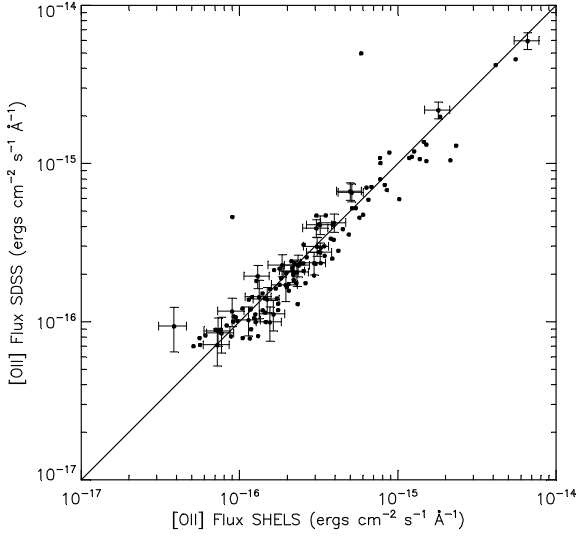}
\caption{Comparison of
[OII] $\lambda$3727 line fluxes for 128 galaxies observed by SHELS and
SDSS. We include those measurements where the EW [OII] $\geq$3 {\rm
\AA} and $\geq$3 $\sigma$ significance. The ratio of the median SDSS
and SHELS measurements is 0.96.  \label{fig9}}
\end{figure}

\begin{figure}
\plotone{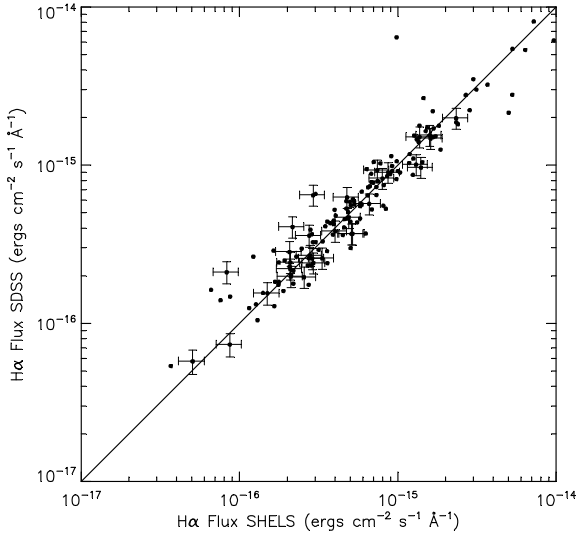}
\caption{Comparison of
H${\alpha}$ line fluxes for 154 galaxies observed by SHELS and
SDSS. We include those measurements where the EW [OII] $\geq$3 {\rm
\AA} and $\geq$3 $\sigma$ significance. The ratio of the median SDSS
and SHELS measurements is 1.03.  \label{fig10}}
\end{figure}

\begin{figure}
\plotone{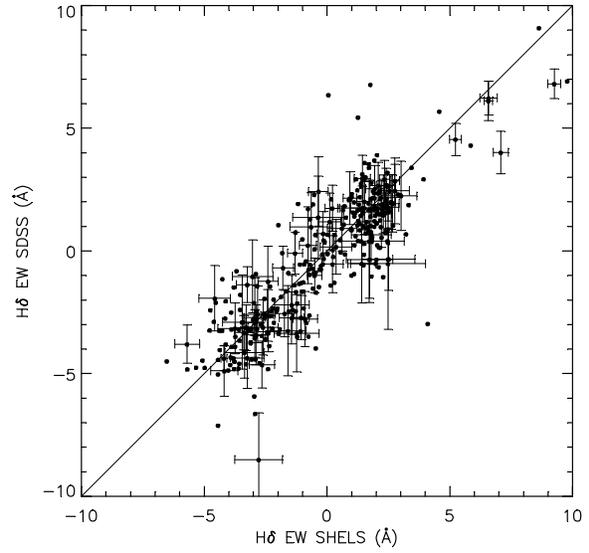}
\caption{Comparison of
H${\delta}$ equivalent widths for 352 galaxies observed by SHELS and
SDSS.  Those galaxies with equivalent width measurement errors $\leq$3
{\rm \AA} are included. The ratio of the median SDSS
and SHELS measurements is 0.97. \label{fig11}}
\end{figure}

\begin{figure}
\plotone{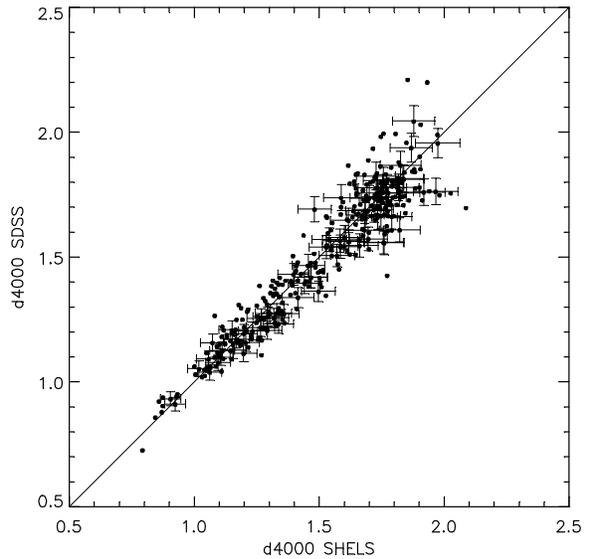}
\caption{Comparison of d4000 for 358 galaxies observed by SHELS and SDSS.  The ratio of the median SDSS and SHELS measurements is 1.00.
\label{fig12}}
\end{figure}

\begin{figure}
\plotone{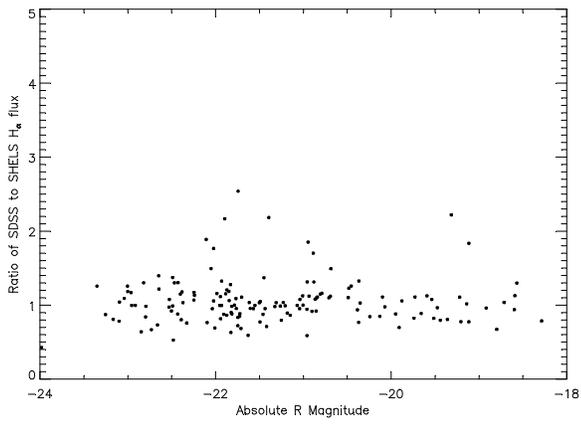}
\caption{Ratio of SDSS to SHELS H$\alpha$ fluxes as a function of R absolute magnitude. No systematic trend is observed, even though the SHELS fibers generally intercept less than 20\% of the light from these galaxies. \label{fig13}}
\end{figure}

\clearpage


\begin{deluxetable}{cccc}
\tablecaption{Definition of Spectral Indices (All given in {\rm \AA})}
\tablehead{
\colhead{Index} & \colhead{Blue Cont. Center (Width)} & \colhead{Feature Center (Width)} &
\colhead{Red Cont. Center (Width)}
}
\startdata
[OII]            & {3685.0 (25.0)} & {3727.3  (8.0)} & {3770.0 (25.0)}\\
{H${\delta}$ A} & {4060.7 (19.1)} & {4102.9 (19.4)} & {4144.8 (16.3)}\\
H${\alpha}$     & {6510.0 (25.0)} & {6562.8  (8.0)} & {6622.0 (25.0)}\\
\enddata
\end{deluxetable}

\begin{deluxetable}{rrll}
\tablecaption{ADC Prism Malfunction Intervals}
\tablehead{
\colhead{Year} & \colhead{Interval} &\colhead{Nature of Error} &
\colhead{Effect on Data}
}
\startdata
2004 & before  10/30 & ADC prisms rotating in wrong sense & serious\\
2005 & 10/25         & ADC prisms off, no correction & small for low airmass\\
2006 & 11/16 to 11/22& ADC prisms off, unknown orientation & unknown\\
2007 & 2/21  to 3/16 & ADC prisms off, unknown orientation & unknown\\
\enddata
\end{deluxetable}

\begin{deluxetable}{lllll}
\tabletypesize{\footnotesize}
\tablecaption{SHELS Spectrophotometry and Spectral Index Internal Error Estimates}
\tablehead{
\colhead{Feature} &
\colhead{Number of Pairs} &
\colhead{SHELS Systematic Error} &
\colhead{Total RMS Scatter} &
\colhead{Fraction $>$2$\sigma$}
}
\startdata
[OII] line flux     & 762 & 18\% of flux& 22\% & 6.8\% \\
H$\alpha$ line flux & 592 & 18\% of flux& 23\% & 9.1\% \\
H$\delta$ EW & 1011 &total error is 2.3 & 1.5 {\rm \AA} & 4.4\% \\
                    &      &times statistical error &       & \\
d4000               & 1468 & total error is 0.045  & 0.086  & 9.8\% \\
                    &      & times value           &        &   \\
\enddata
\end{deluxetable}

\begin{deluxetable}{llllll}
\tabletypesize{\scriptsize}
\tablecaption{Comparison of SHELS and SDSS Spectrophotometry and Spectral Indices}
\tablehead{
\colhead{Feature} &
\colhead{Number of Pairs} &
\colhead{Median Ratio} &
\colhead{SDSS Systematic Error} &
\colhead{Total RMS Scatter} &
\colhead{Fraction $>$2$\sigma$}
}
\startdata
[OII] line flux    & 128 & 0.96 & 12\% of flux& 22\% & 4.7\% \\
H$\alpha$ line flux& 154 & 1.03 & 15\% of flux& 22\% & 9.1\% \\
H$\delta$ EW       & 352 & 0.97 &total error is 1.2 & 1.0 {\rm \AA} & 5.1\% \\
                   &     &      &times statistical error    &      \\
d4000              & 358 & 1.00 &total error is 0.03 & 0.064& 5.6\% \\
                   &     &      &times value         &      &       \\
\enddata
\end{deluxetable}

\end{document}